# TOWARD A RELIABLE GIGABIT NETWORK - AN UPGRADE OF THE SPRING-8 NETWORK

T. Fukui, M. Ishii, R. Tanaka, T. Masuda, T. Ohata
SPring-8, Hyogo 679-5198, Japan

*Abstract*

The SPring-8 controls network has maintained accelerator operations in high reliability and shown good performance during the past years. To cope with the increase of loads on the network due to faster data acquisition and the addition of equipment data, networking hardware has been installed in the last few years. The upgraded network replaces the original FDDI backbone and switches with mixed FDDI/gigabit ethernet and Layer-3 switches. It is necessary to keep the double ring topology for the FDDI and introduce link aggregation technology for the gigabit ethernet to maintain the full redundancy and bandwidth of the system. This paper will discuss the network performance of the gigabit ethernet including its latency and redundancy. We also discuss a future plan for the network including Quality-of-Service over the gigabit ethernet.

## 1 INTRODUCTION

At the beginning of the SPring-8, the network system was constructed from three accelerator controls networks (storage ring, synchrotron, linac), insertion devices (ID-LAN) and beamline controls network (BL-LAN)[1]. The SPring-8 control system has been implemented using the ONC Remote Procedure Call (ONC/RPC) with TCP/IP to connect the workstations and the equipment controllers[2]. Any of the devices on the networks can be controlled from the central control room (CCR).

In the winter of 1999, a new accelerator, a 1.5GeV storage ring named New SUBARU [3], was scheduled to start operation and its control network was connected to the backbone.

In the winter of 2000, the routers between the FDDI backbone and BL-LAN (and ID-LAN) were replaced with Layer-3 switches. In addition, the gigabit ethernet (GE) was installed to serve wider bandwidth for data acquisition systems (DAQ). We considered two things when upgrading the network. The first thing was the rapid increase of beamlines. In the original network several beamlines shared the network at a transfer rate of 10Mbits/sec and the increase in the number of beamlines became a bottleneck in the network traffic. Secondly, we need to provide an infrastructure that can handle the demands of the higher speed DAQs of the accelerator controls.

## 2 NETWORK OVERVIEW

The SPring-8 controls network is shown in Figure 1. The FDDI is used as the backbone of the equipment control VMEs because the double ring topology is redundant. The FDDI switches connect three FDDI backbones of the accelerators and each accelerator control can use the maximum bandwidth of the FDDI (100Mbits/sec). The BL-LAN and ID-LAN are connected to the FDDI backbone via the Layer-3 switches to keep enough bandwidth for these controls. The DAQ networks are connected to the fast ethernet (FE) switches and these are connected to the GE switch on the CCR. The traffic of the DAQs is independent from the equipment controls network. No packet is lost from the equipment controls even if the DAQ sends large amounts of data.

The operator consoles and servers are connected to two FE switches and these are connected to the GE switch. The FDDI switch at the CCR is also connected to the GE switch. The network traffic between the operator consoles and servers and the VMEs are all passed through the GE switch. A bandwidth of this configuration could reach up to the backplane bandwidth of the GE switch (12Gbits/sec).

## 3 PERFORMANCE OF GIGABIT ETHERNET

### 3.1 Setup of Performance Test Bench

Figure 2 shows the test bench setup. One of the PCs running Linux was used to simulate the data acquisition system and another the console. This setup is capable of testing up to three GE switch configurations and to aggregate three links between a Layer-3 switch, CoreBuilder 3500, and a GE switch, SuperStack II 9300.

We used two types of performance test program. One is called Netperf and is developed by Hewlett-Packard and another was a program compiled with ONC/RPC to simulate the SPring-8 control framework.

We measured the bandwidth of the FE and GE through the GE switch using Netperf. The GE shows 360Mbits/sec including the software overhead and the bandwidth of the PC.

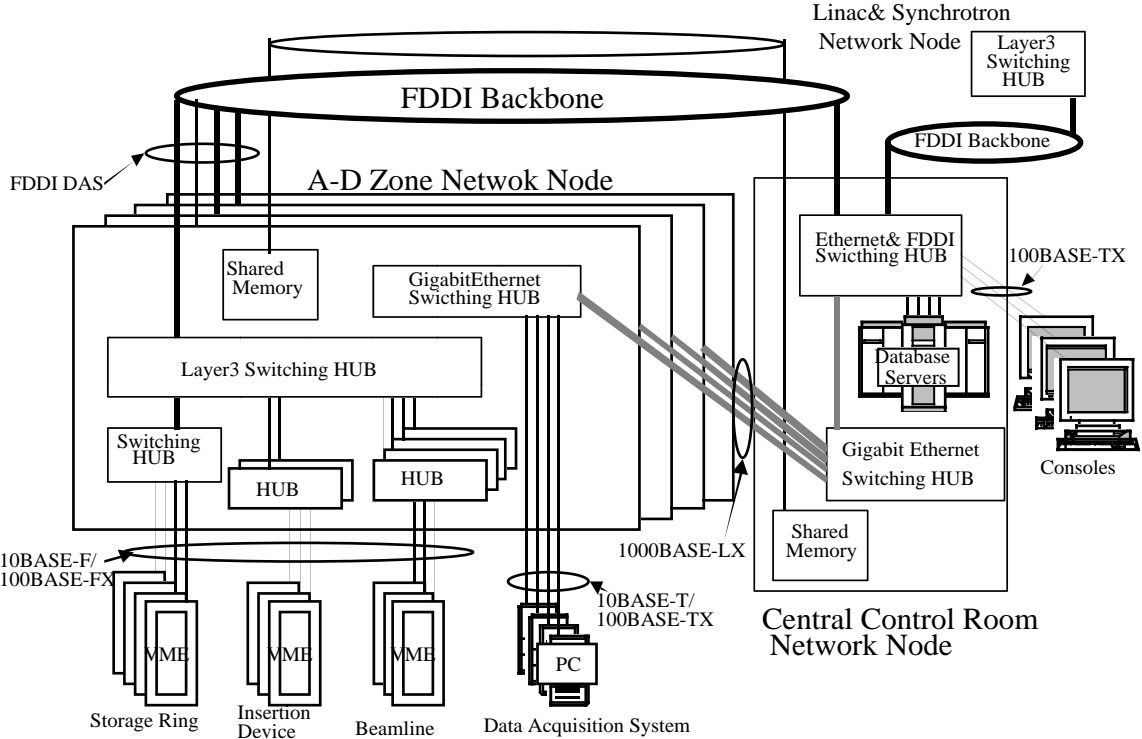

Figure 1:Schematic view of the SPring-8 network.

It is almost 4 times wider bandwidth than the FE. There is no degradation of the bandwidth caused by the GE switch.

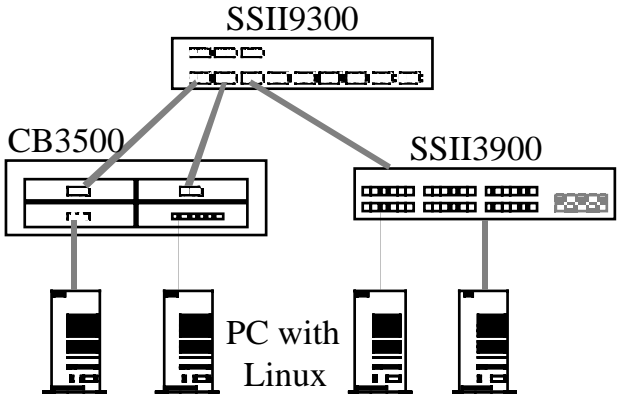

Figure 2:GE test setup

## 3.2 Latency of Gigabit Ethernet Switch

We measured the response time between two PCs to calculate the latency using Netperf. Figure 3 shows the latency of the TCP/IP connection with a FE switch, a GE switch (direct), two GE switches (1 hop) and three GE switches (2 hops). This shows about 10µsec delay with 65bytes (64bytes header + 1byte data) transmission per hop for the GE switch. FE shows a much faster response than GE. This may depend on the device driver for the network interface card but we did not test another GE card.

## 3.3 Latency Degradation of Link Aggregation

The link aggregation (LA) is used to group multiple ports to one logical high-speed link. If several clients are talking to several servers, the load balancing of LA works fine. In this case the total bandwidth with LA is equal to the sum of a port bandwidth. On the other hand any link aggregation must pay a performance penalty because of the overhead for the aggregated port control sequence. The latency of two links aggregate and three links aggregate was measured. The performance degradation of the link aggregation is about 1µsec with no dependence on data size and no difference was observed between two and three links.

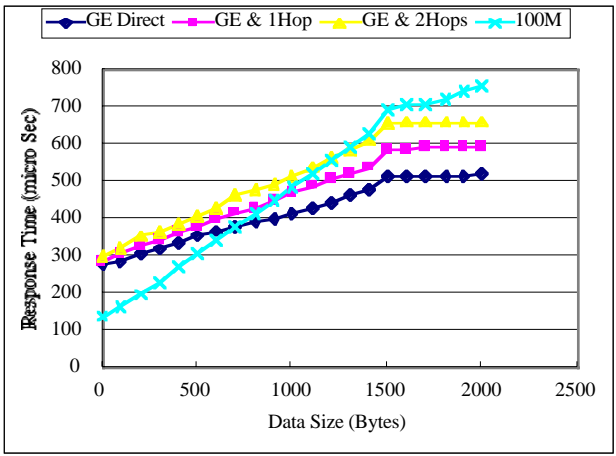

Figure 3: Measured latency of the GE and FE.

### 3.4 Redundancy of Link Aggregation

Spanning Tree Protocol (STP) is probably the oldest and mostly used in the ethernet to keep redundancy. STP usually has a reconnection time between 30 seconds to few minutes. Link aggregation is an upcoming standard and can perform as a redundant link. If one port fails, the transmission is moved to another port. Reconnection time was tested with a 256byte transmission via ONC/RPC. The maximum time for reconnection was about 30 seconds and the median value about three seconds. This is much better than STP but leaves no margin for a transmission timeout of the ONC/RPC default value (25 seconds).

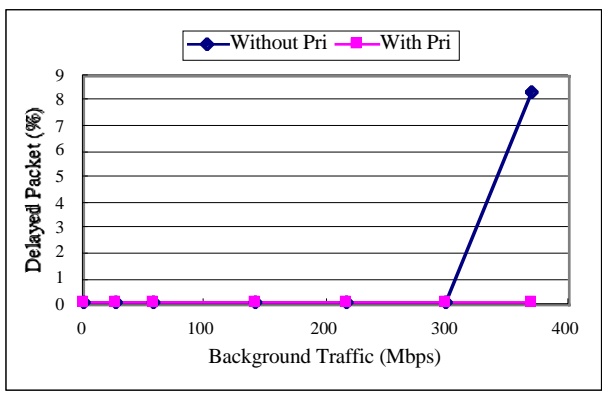

Figure 4: Quality of network with or without priority control.

### 3.5 Quality of Service

The priority control for packets will need to maintain a good quality of service for the accelerator control if the DAQ and the equipment control traffic are carried on the same link. We tested VLAN based priority control. This uses the Type of Service field in the header of the ethernet packet. Four types of service are predefined, those are; Background, Best Effort, Business Critical and Controlled. We set the Controlled class for ONC/RPC to simulate the accelerator control and the Best Effort class for background traffic. Figure 4 shows the percentage of a delayed packet with or without priority control. If we do not use priority control, network utilization must be less than 30 percent to keep a good quality of service for the accelerator control. On the other hand, we can maintain good quality at more than 30 percent if we use priority control. Using priority control necessitates more study to optimize the classes for each type of traffic.

## 4 CONCLUSIONS AND PLAN

We tested the performance of the gigabit ethernet switches. The latency was about 10µsec per hop and 1 µsec increase for the link aggregation configuration. The link aggregation can be used as a redundant link and it has a better reconnection time than STP. Priority control showed a good capability to maintain the quality of service for network traffic.

We have a plan to replace the FDDI with gigabit ethernet. We expect the maintenance cost of the FDDI to increase and that it will be difficult. On the other hand, gigabit ethernet with link aggregation showed good performance and good reliability. We will be able to use the priority control of packet to maintain good quality of service for the accelerator control when utilization becomes more than 30%.

### REFERENCES

[1] R. Tanaka *et al*., "The first operation of control system of SPring-8 storage ring", ICALEPCS'97, Beijing, China, 1997,p.1.
[2] T. Fukui *et al*., "Design and performance of the network system for the storage ring control at SPring-8", ICALEPCS'97, Beijing, China, 1997,p312.
[3] A. Ando *et al*.,"VUV and Soft X-ray Light Source NEW SUBARU", PAC'97, Vancouver, Canada, 1997.